\documentclass[a4paper,11pt]{article}
\pdfoutput=1 

\usepackage{jcappub}
\usepackage{dsfont}
\usepackage{dcolumn}

\title{Dynamics of the diffusive DM-DE interaction--dynamical system approach}

\author[a]{Zbigniew Haba}
\author[b]{Aleksander Stachowski}
\author[b,c]{Marek Szyd{\l}owski}

\affiliation[a]{Institute of Theoretical Physics, University of Wroclaw, \\Plac Maxa Borna 9, 50-204 Wroc{\l}aw, Poland}
\affiliation[b]{Astronomical Observatory, Jagiellonian University, Orla 171, 30-244 Krakow, Poland}
\affiliation[c]{Mark Kac Complex Systems Research Centre, Jagiellonian University, {\L}ojasiewicza 11, 30-348 Krak{\'o}w, Poland}

\emailAdd{zhab@ift.uni.wroc.pl}
\emailAdd{aleksander.stachowski@uj.edu.pl}
\emailAdd{marek.szydlowski@uj.edu.pl}

\abstract{We discuss dynamics of a model of an energy transfer between dark energy (DE) and dark matter (DM). The energy transfer is determined by a non-conservation law resulting from a diffusion of dark matter in an environment of dark energy. The relativistic invariance defines the diffusion in a unique way. The system can contain baryonic matter and radiation which do not interact with the dark sector. We treat the Friedman equation and the conservation laws as a closed dynamical system. The dynamics of the model is examined using the dynamical systems methods for demonstration how solutions depend on initial conditions. We also fit the model parameters using astronomical observation: SNIa, $H(z)$, BAO and Alcock-Paczynski test. We show that the model with diffuse DM-DE is consistent with the data.
}

\begin{document}
\maketitle
\flushbottom

\section{Introduction}

In spite of an excellent agreement of the $\Lambda$CDM model
with observational data  some basic assumptions of this model need
justification. There are some ingredients in the model
which could hardly be derived from a certain fundamental theory.
The presence of dark energy (DE)with its currently small value
is difficult to explain in the standard model of elementary particles \cite{Weinberg:1988cp}.
Then, the relation of dark energy to the dark matter (DM) seems accidental
(coincidence problem). That these components are of the same order suggests
that there may be certain dynamical relation between them.
We suggest a model describing an irreversible flow of DE to DM. We
assume that the total mass of the dark matter does not change.
These assumptions lead to the unique model of the DM-DE
interaction.

The Einstein equations are
\begin{equation} R^{\mu\nu}-\frac{1}{2}g^{\mu\nu}R=T^{\mu\nu},\label{eq:1}
\end{equation}
where $R^{\mu\nu}$ is the Ricci tensor, $g^{\mu\nu}$ the metric and $8\pi G = c = \hbar = 1$, we can decompose the right-hand sides of (\ref{eq:1}) as
\begin{equation}
T^{\mu\nu}=T^{\mu\nu}_{b}+T^{\mu\nu}_{R}+T_{de}^{\mu\nu}+T_{dm}^{\mu\nu},\label{eq:2}
\end{equation}
where the absence of an interaction between baryonic matter
$T_{b}$, radiation $T_{R}$ and the dark component means
\begin{equation}
\nabla_{\mu}(T^{\mu\nu}_{R}+T^{\mu\nu}_{b})=0.\label{eq:3}
\end{equation}
The conservation of the total energy gives
\begin{equation}\label{eq:4}
\nabla_{\mu}T_{\text{de}}^{\mu\nu}=-\nabla_{\mu}T_{\text{dm}}^{\mu\nu}\equiv -
3\kappa^{2} J^{\nu}
\end{equation}
with a current $J^{\nu}$ and a certain constant $\kappa$ which can
be calculated when the model of $T_{dm}^{\mu\nu}$ is defined.

The relation between the non-conservation law (\ref{eq:4}) can explain the
coincidence between DM and DE densities as well as the relevance
of the dark energy exactly at the present epoch. We need a model
for $T_{de}^{\mu\nu}$ and $T_{dm}^{\mu\nu}$. We assume that the
gain of energy of the dark matter consisting of particles of mass
$m$ results from a diffusion in an environment described by an
ideal fluid. There
 is only one diffusion which is
relativistic invariant and preserves the particle mass $m$
\cite{Dudley:1965lm}. The corresponding energy-momentum satisfies the
conservation law (\ref{eq:4}). The current $J^{\nu}$ in eq.~(\ref{eq:4}) is
conserved \cite{Haba:2010qj,Calogero:2011re,Calogero:2012kd}
\begin{equation}
\nabla_{\mu}J^{\mu}=0.\label{eq:5}
\end{equation}
This is a realization of the conservation law of the total mass of the dark
matter. In a homogeneous universe the current conservation implies
\begin{equation}
J^{0}=\frac{\gamma}{3\kappa^{2}} a^{-3},\label{eq:6}
\end{equation}
where $a$ is the scale factor of an expanding metric and another constant $\gamma$.

In a homogeneous space-time we can represent the DM as well as DE
energy-momentum as the energy-momentum of an ideal fluid. The
conservation law (\ref{eq:4}) leads to a particular interaction among the
fluids. An interaction which is a linear combination of the DM and
DE fluids has been discussed in\cite{Zimdahl:2014jsa}. Non-linear
interactions are discussed in \cite{Perez:2013zya,Boehmer:2008av,Bolotin:2013jpa,Boehmer:2014vea}. Our formula for the DM dissipation (\ref{eq:4})
follows from the assumption that the dissipation results from a
relativistic motion in a DE fluid. It cannot be expressed as a
polynomial formula in DM and DE fluids as it is in the above
mentioned references. Nevertheless, we are able to express the
dynamics of the model as a quadratic dynamical system what makes
our approach similar to that of refs.~\cite{Perez:2013zya,Boehmer:2008av,Boehmer:2014vea}.

Methods of dynamical systems \cite{Perko:2001de} have been recently used in
a cosmological model with diffusion described by a cosmological
scalar field \cite{Alho:2014ola}. A similar analysis of the
dynamics has been also explored in the context of Bianchi
cosmological models \cite{Shogin:2014taa} as well as in a
description of non-homogeneous and anisotropic cosmological models
\cite{Shogin:2013nra}. In this paper we intend to explore
cosmological models as closed dynamical systems with matter (dark
and baryonic) and dark energy in the form of ideal fluids whose
interaction is determined by the current $J^{\nu}$ (\ref{eq:4}). In
contradistinction to above mentioned models our model does not
contain non-physical trajectories passing through
$\rho_{\text{m}}=0$ line \cite{Stachowski:2016dfi}.

The plan of the paper is the following. In sec.~2 we review the
model of a relativistic diffusion and explain eq.~(\ref{eq:4}). In sec.~3 we
derive exactly soluble limits relevant for early and late
universe. We discuss energy-momentum conservation and Einstein
equations in sec.~4. In sec.~5 we formulate the cosmological
equations of sec.~4 as a closed dynamical system. We determine its
critical points and the phase portrait. In sec.~6 we fit the
parameters of the model to the observational data.

\section{Relativistic diffusion}

In this section we consider a Markovian
approximation of an interaction of the system with an environment
which leads to the description of this interaction by a diffusion.
We consider a relativistic generalization of the Krammers
diffusion defined on the phase space. It is determined in the
unique way by the requirement that the diffusing particle moves on
the mass-shell (see \cite{Dudley:1965lm,Franchi:2007rd,Haba:2008uy,Haba:2009by}).

Let us choose the contravariant spatial coordinates $p^{j}$ on the
mass shell and define the Riemannian metric
\begin{displaymath}
ds^{2}=g_{\mu\nu}dp^{\mu}dp^{\nu}=-G_{jk}dp^{j}dp^{k},
\end{displaymath}
where Greek indices $\mu,\nu=0,1,2,3$, Latin indices $j,k=,1,2,3$
and $p_{0}$ is expressed by $p^{j}$ from $p^{2}=m^{2}$. We have
(we assumed that $ g_{0k}=0$)
\begin{displaymath}
G_{jk}=-g_{jk}+p_{j}p_{k}\omega^{-2},
\end{displaymath}
where
\begin{displaymath}
\omega^{2}=m^{2}-g_{jk}p^{j}p^{k}.
\end{displaymath}
Then, the inverse matrix is
\begin{displaymath}
G^{jk}=-g^{jk}+m^{-2}p^{j}p^{k}.
\end{displaymath}
Next,
\begin{displaymath}
G\equiv -\det(G_{jk})=-m^2\det(g_{jk})\omega^{-2}.
\end{displaymath}

We define diffusion as a stochastic process generated by
the Laplace-Beltrami operator $\triangle_{H}^{m}$ on the mass shell
\begin{equation}
\triangle_{H}^{m}=\frac{1}{\sqrt{G}}\partial_{j}G^{jk}\sqrt{G}\partial_{k},\label{eq:7}
\end{equation}
where $\partial_{j}=\frac{\partial}{\partial p^{ j}}$ and
$G=\det(G_{jk})$ is the determinant of $G_{jk}$.

The transport equation for the diffusion generated by
$\triangle_{H}$ reads
\begin{equation}
\begin{array}{l}
(p^{\mu}\partial^{x}_{\mu}-\Gamma^{k}_{\mu\nu}p^{\mu}p^{\nu}\partial_{k})\Omega=
\kappa^{2}\triangle^{m}_{H}\Omega,\label{eq:8}
\end{array}\end{equation}
where $\Gamma^{k}_{\mu\nu}$ are the Christoffel symbols,
$\partial_{\mu}^{x}$ are space-time derivatives and $\kappa^{2}$
is the diffusion constant.

Then, we can define the current
\begin{equation}
\begin{array}{l} J^{\mu}=\sqrt{g}\int \frac{d{\bf
p}}{(2\pi)^{3}}p_{0}^{-1}p^{\mu}\Omega
\end{array},\label{eq:9}
\end{equation}
where $g=\vert\det [g_{\mu\nu}]\vert$ and the energy momentum

\begin{equation}\begin{array}{l}
T_{de}^{\mu\nu}=\sqrt{g}\int  \frac{d{\bf
p}}{(2\pi)^{3}}p_{0}^{-1}p^{\mu}p^{\nu}\Omega
\end{array}.\label{eq:10}
\end{equation}
It can be shown using eq.~(\ref{eq:8}) that \cite{Haba:2010qj,Calogero:2011re,Calogero:2012kd}
\begin{equation}
\nabla_{\mu}T_{\text{dm}}^{\mu\nu}=3\kappa^{2}J^{\nu}\label{eq:11}
\end{equation} and

\begin{equation}
\nabla_{\mu}J^{\mu}=g^{-\frac{1}{2}}\partial_{\mu}(g^{\frac{1}{2}}J^{\mu})=0.\label{eq:12}
\end{equation}
Hence,
\begin{equation}
 g^{-\frac{1}{2}}\partial_{t}( g^{\frac{1}{2}}J^{0})=-\partial_{j}J^{j}.\label{eq:13}
\end{equation}
This implies (\ref{eq:6}) if the metric is homogeneous and $\Omega $ does
not depend  on $x$.
 The
constant $\gamma$ can be expressed from eq.~(\ref{eq:9}) as
\begin{equation}\frac{\gamma}{3\kappa^{2}}
=g\int  \frac{d{\bf p}}{(2\pi)^{3}}\Omega\equiv Z.\label{eq:14}
\end{equation}

\section{The limits $ma\rightarrow 0$ and $ma\rightarrow \infty$}

Most of our subsequent results hold true for a general FWR metric
\begin{equation}
ds^{2}=g_{\mu\nu}dx^{\mu}dx^{\nu}=dt^{2}-a^{2}h_{jk}dx^{j}dx^{k},\label{eq:15}
\end{equation}
but for simplicity of our analysis we restrict ourselves to the
flat space $h_{jk}=\delta_{jk}$. We rewrite the diffusion equation
in terms of the covariant momenta
\begin{equation}
q_{j}=g_{ij}p^{j}\label{eq:16}
\end{equation}
Then, $\triangle_{H}^{m}$ in eq.~(\ref{eq:7}) depends on $
\sqrt{m^{2}a^{2} +{\bf q}^{2}}$ and on $a$. The assumption ${\bf
	q}^{2} \gg m^{2}a^{2}$ (high energy approximation) is equivalent to the limit
\begin{equation}
m^{2}a^{2}\rightarrow 0.\label{eq:17}
\end{equation}
Let 
\begin{equation} 
\nu=\int dt \, a
\end{equation}
Then, in the limit $m^{2}a^{2}\rightarrow 0$
and in a homogeneous universe ($\Omega$ independent of spatial coordinates) we obtain
\begin{equation}
\kappa^{-2}\vert {\bf q}\vert
\partial_{\nu}\Omega=q_{i}q_{j}\frac{\partial^{2}}{\partial q_{i}\partial q_{j}}\Omega
+3q_{j}\frac{\partial}{\partial q_{j}}\Omega\label{eq:18}
\end{equation}or in the original contravariant coordinates
\begin{equation}
a\kappa^{-2}\vert {\bf p}\vert
(\partial_{t}-2Hp^{j}\frac{\partial}{\partial
	p^{j}})\Omega=p^{i}p^{j}\frac{\partial^{2}}{\partial p^{i}\partial
	p^{j}}\Omega +3p^{j}\frac{\partial}{\partial p^{j}}.\Omega\label{eq:19}
\end{equation}
If in 
$\sqrt{m^{2}a^{2} +{\bf q}^{2}}$ we assume  ${\bf q}^{2} \ll
m^{2}a^{2}$ (low energy approximation,i.e.,
we neglect ${\bf q}$)
then in the limit
\begin{displaymath}
m^{2}a^{2}\rightarrow \infty
\end{displaymath}
eq.~(\ref{eq:8})  simplifies to
\begin{equation}
m^{-1}\kappa^{-2}\partial_{\sigma}\Omega=\frac{1}{2}\triangle_{\bf
	q}\Omega,\label{eq:20}
\end{equation}
where
\begin{equation} \sigma=2\int_{t_{0}}^{t} ds \, a^{2}\label{eq:21}
\end{equation}
and $\triangle_{\bf q}$ is the Laplacian.
This is the non-relativistic diffusion equation. In terms of the
original contravariant momenta eq.~(\ref{eq:20}) takes the form
\begin{equation}
m^{-1}a^{2}\kappa^{-2}\left(\partial_{t}-2Hp^{j}\frac{\partial}{\partial
	p^{j}} \right)\Omega=\frac{1}{2}\triangle_{\bf p}\Omega.\label{eq:22}
\end{equation}

\section{Current conservation and Einstein equations}

The energy-momentum (\ref{eq:10}) in a homogeneous space-time can be
expressed as an energy-momentum of a fluid
\begin{equation}
T^{\mu\nu}=(\rho+p)u^{\mu}u^{\nu}-g^{\mu\nu}p,\label{eq:23}
\end{equation}where \begin{equation}
g_{\mu\nu}u^{\mu}u^{\nu}=1.\label{eq:24}\end{equation}

The divergence equations (\ref{eq:4}) in the frame $u=(1,{\bf 0})$ takes the form
\begin{equation}
\partial_{t}\rho_{dm}+3H(1+\tilde{w})\rho_{dm}=\gamma a^{-3},\label{eq:25}
\end{equation}where
\begin{displaymath}
\tilde{w}=\frac{p_{dm}}{\rho_{dm}}.
\end{displaymath}
We assume that the energy-momentum tensor of the dark energy has
also the form of an ideal fluid~(\ref{eq:23}). Then, from eqs.~(\ref{eq:4}) and (\ref{eq:6})
\begin{equation}
\partial_{t}\rho_{de}+3H(1+w)\rho_{de}=-\gamma a^{-3}.\label{eq:26}
\end{equation}
On the basis of
observational data we choose $w=-1$ in eq.~(\ref{eq:26}). In the diffusion
model $\tilde{w}$ depends on time as follows from the formula
\begin{displaymath}\begin{array}{l}
\tilde{w}=\frac{1}{3}\int d{\bf p}\frac{1}{p_{0}}a^{2}{\bf
	p}^{2}\Omega_{t}\Big(\int d{\bf p}p_{0}\Omega_{t}\Big)^{-1}\cr
=\frac{1}{3}-\frac{m^{2}}{3}\int d{\bf
	p}\frac{1}{p_{0}}\Omega_{t}\Big(\int d{\bf
	p}p_{0}
\Omega_{t}\Big)^{-1}\equiv \frac{1-\omega}{3}.
\end{array}\end{displaymath}
In an expansion in $m$ we can apply the explicit solution
\cite{Haba:2013xka} of eq.~(\ref{eq:19}) ($m=0$) and calculate
\begin{displaymath}
\omega=\frac{m^{2}a^{2}}{6(T_{0}+\kappa^{2} \nu)},
\end{displaymath}
where $T_{0}$ is a parameter which has the meaning of the
temperature of the DM fluid at $t=t_{0}$. In the
ultrarelativistic (massless) case (\ref{eq:19}) we have $\omega=0$, hence
$\tilde{w}=\frac{1}{3}$.

We can express the solution of eq.~(\ref{eq:25}) as
\begin{multline}
\rho_{dm}(t)=\rho_{dm}(0)a^{-4}\exp\left(\int_{t_{0}}^{t}d\tau \,
H\omega\right) \\
+\gamma a^{-4}\exp\left(\int_{t_{0}}^{t}d\tau \,
H\omega\right)\int_{t_{0}}^{t}ds \, a(s)\exp\left(-\int_{t_{0}}^{s}d\tau \,
H\omega\right).\label{eq:27}
\end{multline} For $w=-1$
\begin{equation}
\rho_{de}(t)=\rho_{de}(0)-\gamma \int_{t_{0}}^{t}a^{-3 }(s)\, ds.\label{eq:28}
\end{equation}
We still consider the non-relativistic limit of the energy-momentum (\ref{eq:10})
\begin{multline}
\rho_{\text{dm}}=\tilde{T}^{00}=\sqrt{g}(2\pi)^{-3}\int d\mathbf{
p} \, p^{0}\Omega =g^{-\frac{1}{2}}Zm+\sqrt{g}(2\pi)^{-3}\int d{\bf
p} \, \frac{a^{2}{\bf p}^{2}}{2m}\Omega\cr\equiv
Zma^{-3}+a^{-2}\rho_{nr},\label{eq:29}
\end{multline}
where \begin{equation}\rho_{nr}=\sqrt{g}(2\pi)^{-3}\int d{\bf
p}\, \Omega a^{4}\frac{{\bf p}^{2}}{2m}.\label{eq:30}
\end{equation}

Using the non-relativistic diffusion equation (\ref{eq:22}) we can show
that $\rho_{nr}$ satisfies the non-conservation equation
\begin{equation}
\partial_{t}a^{-2}\rho_{nr}+
3H(1+\tilde{w}_{nr})a^{-2}\rho_{nr}=\gamma a^{-3},\label{eq:31}
\end{equation}
where $\tilde{w}_{nr}=\frac{2}{3}$. The non-relativistic diffusive energy in eq.~(\ref{eq:29}) is a sum of two terms
$Zma^{-3}$ which describes a conservative non-relativistic total
rest mass and $a^{-2}\rho_{nr}$ describing the diffusive energy
gained from the motion in an environment of the dark energy. The
sum of these energies satisfies the equation (which is not of the
form) (\ref{eq:4}))
\begin{equation}
\partial_{t}\rho_{\text{dm}}+5H\rho_{\text{dm}}=3Z\kappa^{2}
a^{-3}+2ZmH a^{-3}.\label{eq:32}
\end{equation}

The solution of eq.~(\ref{eq:31}) is
\begin{equation}
\tilde{\rho}_{nr}(t)=a^{-3}(\rho_{nr}(0)+\frac{1}{2}\gamma\sigma),\label{eq:33}
\end{equation}
where $\sigma$ is defined in eq.~(\ref{eq:21}). As a consequence of
eqs.~(\ref{eq:29}), (\ref{eq:31}) and (\ref{eq:4}) the non-relativistic dark energy satisfies
the same eqs.~(\ref{eq:26}) and (\ref{eq:28}) as the relativistic dark energy.

The Friedman equation in the FRW metric~(\ref{eq:15}) with the dark matter, dark energy and  baryonic matter $\rho_{b}$ reads
\begin{equation}
\begin{array}{l}
H^{2}=\frac{1}{3}(\rho_{\text{dm}}+\rho_{\text{de}}+\rho_{b}).
\end{array}\label{eq:34}
\end{equation}
By differentiation
\begin{equation}
\dot{H} =-\frac{1}{2}\Big((1+\tilde{w})\rho_{\text{dm}}+\rho_{b}\Big).\label{eq:35}
\end{equation}
In eqs.~(\ref{eq:34})-(\ref{eq:35}) we should insert the general expressions for DM
and DE. We need an
approximation for $\tilde{w}(t)$. There we shall discuss
approximations to eq.~(\ref{eq:27}). In a subsequent section we study the
relativistic homogeneous dynamical system (\ref{eq:25}), (\ref{eq:26}) and (\ref{eq:35}) under
the assumption that $\tilde{w}$ is time independent. The
non-relativistic (low $z$) approximation (\ref{eq:29}) when inserted in
eq.~(\ref{eq:34}) gives the Friedmann equation
\begin{multline}
H^{2} =\frac{1}{3}\Big(a^{-5}(\rho_{dm}(0)+\frac{3}{2}Z\kappa^{2}\sigma)+Zma^{-3}+\rho_{b}(0)a^{-3} \\
+\rho_{de}(0)-3Z\kappa^{2} \int_{t_{0}}^{t} ds \, a(s)^{-3}
\Big).\label{eq:36}
\end{multline}
Eqs.~(\ref{eq:28}), (\ref{eq:32}) and (\ref{eq:36}) form a system of ordinary differential equations which is expressed by means of new (energetic) variables into a quadratic dynamical system in the next section.

\section{Dynamical system approach to the DM-DE interaction}

In this section we reduce the dynamics of the diffusive DM-DE
interaction to the form of autonomous dynamical system
$\frac{dx}{dt} = \dot{x}=f(x)$, where $x$ is a state variable and
$t$ is time. In this approach one describes the evolution of the
diffusive DM-DE interaction in terms of trajectories situated in a
space of all states of the system, i.e., a phase space. This space
possesses the geometric structure which is a visualization of a
global dynamics, i.e. it is the space of all evolutional paths of
the physical system, which are admissible for all initial
conditions. The equivalence of phase portraits is established by
means of a homeomorphism (topological equivalence) which is
mapping trajectories of the system while preserving their
orientation. The phase space is organized by critical points,
which from the physical point of view  represent stationary states
of the system. From the mathematical point of view they are
singular solutions of the system $\dot{\mathbf{x}}=f(\mathbf{x})$,
where $\mathbf{x}\in\mathbb{R}^n$ is a vector state, corresponding
to vanishing right-hand sides of the system, i.e. $\mathbf{f}=[f^1
(x),\dots,f^n (x)]$ and $\forall_i f^i (x)=0$. The final outcome
of any dynamical system analysis is the phase portrait of the
system from which one can easily obtain the information about the
stability and genericity of particular solutions.

The methods of dynamical systems \cite{Perko:2001de}, which enable
us to investigate the dynamics of the system without the knowledge
of its exact solutions, have been recently applied in a similar
context of cosmological models with diffusion \cite{Alho:2014ola}. An
analysis of cosmological dynamics has also been explored in
Bianchi cosmological models \cite{Shogin:2014taa}. Some of these
methods are applicable  to non-homogeneous and anisotropic
cosmological models \cite{Shogin:2013nra} as well. In this paper
we intend to explore an energy exchange in models describing
matter (dark and baryonic) and dark energy in the form of the
cosmological ideal fluids. In contrast to Alho et al. \cite{Alho:2014ola}
our model does not contain non-physical trajectories passing
through $\rho_{\text{m}}=0$ line \cite{Stachowski:2016dfi}.

\subsection{Cosmological models with constant equation of state for DM and cosmological constant---dynamical system analysis}
Let us consider the continuity equations for the model with
$\tilde{w}=\text{const}$ and $w=-1$ (dark energy in the form of
the cosmological constant). The corresponding continuity equations
take the form (\ref{eq:25})-(\ref{eq:26})
\begin{equation}
 a^{-3(\tilde{w}+1)}\frac{d}{dt}(\rho_{\text{dm}} a^{3(\tilde{w}+1)})=\gamma
a^{-3}>0,\label{eq1}
\end{equation}
\begin{equation}
\frac{d\rho_{\text{de}}}{dt}=-\gamma a^{-3}<0,\label{eq2}
\end{equation}
where $\gamma>0$. To formulate the dynamics in the form of a
dynamical system, we rewrite in a suitable way equation
(\ref{eq1})
\begin{equation}
J=\frac{d\rho_{\text{dm}}/\rho_{\text{dm}}}{da/a}\equiv\frac{d\ln
\rho_{\text{dm}}}{d(\ln a)}=-3(1+\tilde{w})+\frac{\gamma
a^{-3}}{H\rho_{\text{dm}}}.\label{eq3}
\end{equation}
Next we define a dimensionless quantity, which measures the
strength of the interaction
\begin{equation}
\delta\equiv\frac{\gamma a^{-3}}{H\rho_{\text{dm}}}.\label{eq4}
\end{equation}
Clearly, in  general $\delta$ is time dependent. Let us consider
that $\delta=\delta(a(t))$. If this quantity is constant during
the cosmic evolution, then the solution of eq.~(\ref{eq3}) has a
simple form
\begin{equation}
\rho_{\text{dm}}=\rho_{\text{dm},0}a^{-3(1+\tilde{w})+\delta}.
\end{equation}
Our aim is to study the dynamics of the energy transfer from the
DE to DM sector. The corresponding system assumes the form of
a three-dimensional dynamical system.

Recalling that $8\pi G=c=1$ we define
\begin{equation}
x\equiv \frac{\rho_{\text{dm}}}{3H^2},\quad
y\equiv\frac{\rho_{\Lambda}}{3H^2},\label{eq6}
\end{equation}
where $H=\frac{d \ln a}{dt}$ is the Hubble parameter and $t$ is
the cosmological time. The differentiation with respect to the
cosmological time $t$ will be denoted by a dot
($\dot{}\equiv\frac{d}{dt}$). The variables $x$ and $y$ have the
meaning of dimensionless density parameters.

For simplicity of presentation it is assumed that FRW space is
flat (zero curvature in the Friedmann equations (\ref{eq:34})) . In this
case the acceleration equation assumes the following form
\begin{equation}
\dot{H}=-\frac{1}{2}(\rho_{\text{eff}}+p_{\text{eff}}),\label{eq7}
\end{equation}
where $\rho_{\text{eff}}=\rho_{\text{dm}}+\rho_{\text{de}}$ and
$p_{\text{eff}}=\tilde{w}\rho_{\text{dm}}-\rho_{\text{de}}$ are the effective energy
density and pressure of the matter filling the universe,
$\tilde{w}=\frac{p_{\text{dm}}}{\rho_{\text{dm}}}$.

Taking a natural logarithm of the state variables (\ref{eq6}) and
the interaction effect variable (\ref{eq4}) and performing the
differentiation with respect to the cosmological time $t$ we
obtain
\begin{align}
\frac{\dot{x}}{x} &= \frac{\dot{\rho}_{\text{dm}}}{\rho_{\text{dm}}}-2\frac{\dot{H}}{H}=-3H(1+\tilde{w})+\delta H-2\frac{\dot{H}}{H},\label{eq8a}\\
\frac{\dot{y}}{y} &= \frac{\dot{\rho}_{\Lambda}}{\rho_{\Lambda}}-2\frac{\dot{H}}{H}=-\delta H\alpha-2\frac{\dot{H}}{H},\label{eq8b}\\
\frac{\dot{\delta}}{\delta} &=
-3H-\frac{\dot{H}}{H}-\frac{\dot{\rho}_{\text{dm}}}{\rho_{\text{dm}}}=3\tilde{w}H-\delta
H-\frac{\dot{H}}{H},\label{eq8c}
\end{align}
where $\alpha=\frac{\rho_{\text{dm}}}{\rho_{\Lambda}}$.

It would be convenient to divide both sides of the
system~(\ref{eq8a})-(\ref{eq8c}) by $H$ and then reparameterize
the original time variable $t$ following the rule
\begin{equation}
t\rightarrow \tau=\ln a. \label{eq9}
\end{equation}
The differentiation with respect to the parameter $\tau$ will be
denoted by a prime (${}'\equiv\frac{d}{d\tau}$). Note that
$\frac{d\tau}{da}=a^{-1}$ is a strictly monotonic function of the
scale factor $a$.

After the time reparameterization (\ref{eq9}) the
system~(\ref{eq8a})-\ref{eq8c} can be expressed as  the
three-dimensional system of equations
\begin{align}
x'&=x\left(-3(1+\tilde{w})+\delta-2\frac{\dot{H}}{H^2}\right),\label{eq10a}\\
y'&=y\left(-\delta\alpha-2\frac{\dot{H}}{H^2}\right),\label{eq10b}\\
\delta'&=\delta\left(3\tilde{w}-\delta-\frac{\dot{H}}{H^2}\right),\label{eq10c}
\end{align}
where $\frac{\dot{H}}{H^2}$ can be determined from the formula
(\ref{eq7})
\begin{equation}
\dot{H}=-\frac{1}{2}(1+\tilde{w})\rho_{\text{dm}}=-\frac{3}{2}(1+\tilde{w})H^2
x,
\end{equation}
i.e.,
\begin{equation}
\frac{\dot{H}}{H^2}=-\frac{3}{2}(1+\tilde{w})x.\label{eq11}
\end{equation}

In this way the dynamics of the process of decaying cold dark
matter satisfying the equation of state
$p_{\text{dm}}=\tilde{w}\rho_{\text{dm}}$ in the background of the
flat  FRW metric can be described by means  of the dynamical
system theory. The resulting  three-dimensional dynamical system
has the form
\begin{align}
x'&=x\left(-3(1+\tilde{w})+\delta+3(1+\tilde{w})x\right),\label{eq12a}\\
y'&=x\left(-\delta+3(1+\tilde{w})y\right),\label{eq12b}\\
\delta'&=\delta\left(3\tilde{w}-\delta+\frac{3}{2}(1+\tilde{w})x\right),\label{eq12c}
\end{align}
where $\alpha=x/y$.

Note that the right-hand sides of the dynamical system
(\ref{eq12a})-(\ref{eq12c}) are of a polynomial form. Therefore
all methods of dynamical system analysis, especially analysis of
the behavior on the Poincar{\'e} sphere, can be adopted; both in a
finite domain as well as at infinity. One can see that the system
(\ref{eq12a})-(\ref{eq12c}) has as an invariant submanifold
$\{\frac{\dot{H}}{H^2}=0\}$,  the set $\{x\colon x=0\}$,
corresponding to the case of the vanishing dark matter energy
density.

Clearly, the system (\ref{eq12a})-(\ref{eq12c}) has also an
invariant submanifold $\{\delta\colon \delta=0\}$ corresponding to
the case of the vanishing interacting term $Q=\delta$ as it
appears in the $\Lambda$CDM model.

Another interesting submanifold is the plane $\{y\colon
y=\frac{\delta}{3(1+\tilde{w})}\}$. 

\subsection{Dynamics of the model for dust matter}

\begin{figure}
    \centering
    \includegraphics[width=0.7\linewidth]{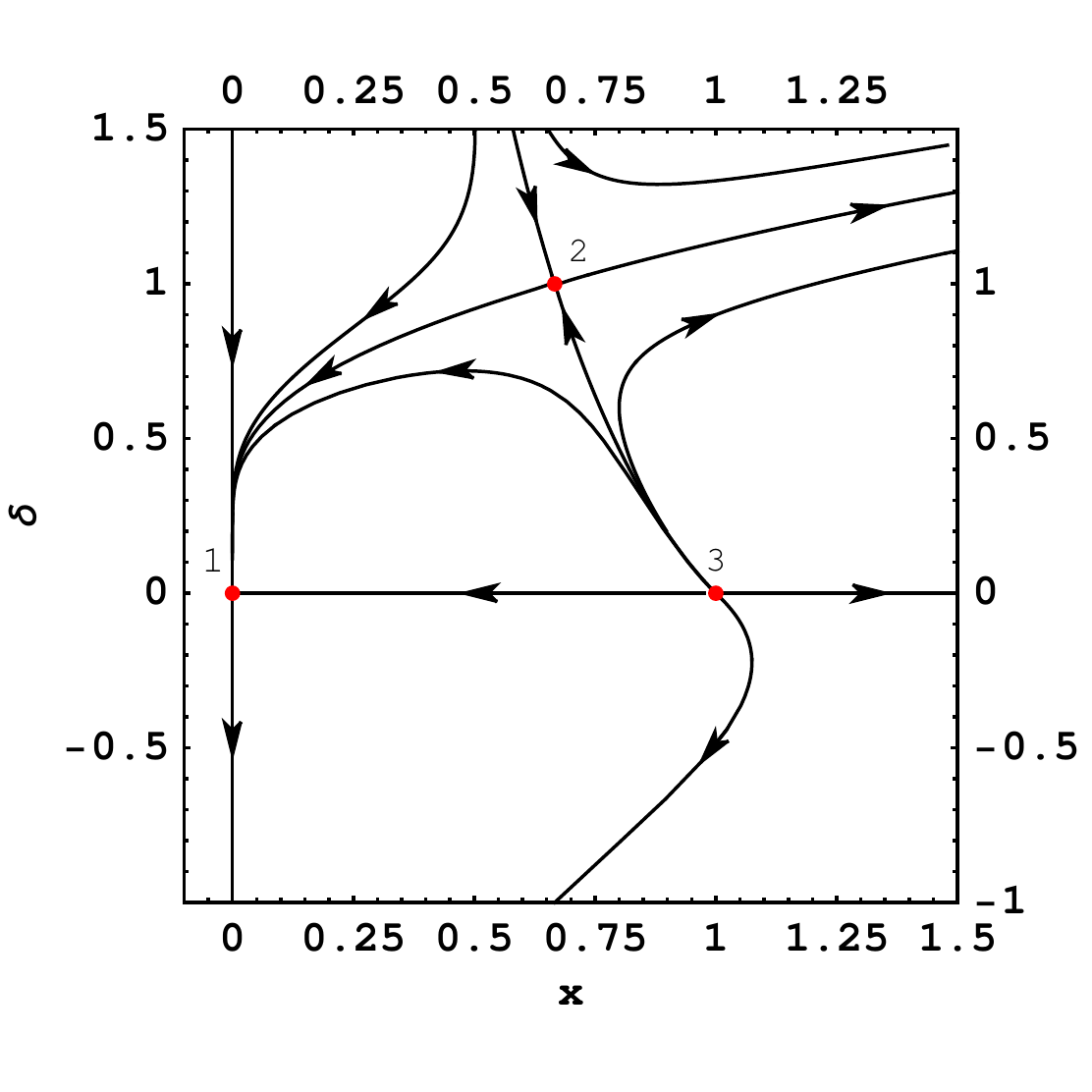}
    \caption{A phase portrait for dynamical system (\ref{eq13a})-(\ref{eq13b}). Critical point (1) ($x=0$, $\delta=0$) represents the de Sitter universe. Critical point (2) ($x=2/3$, $\delta=1$) is a saddle and represents the scaling universe. Critical point (3) ($x=1$, $\delta=0$) is an unstable node and represents the Einstein-de Sitter universe. The critical point (1) is a complex type of saddle-node.}
    \label{fig:2}
\end{figure}

Let $\tilde{w}$ be equal to zero. Then, the equation of state for matter
is of the form of a dust.
Because $x+y=1$ ($\Omega_{\text{dm}}+\Omega_{\text{de}}=1$) then
the dynamical system (\ref{eq12a})-(\ref{eq12c}) reduces to the
two-dimensional dynamical system in the following form
\begin{align}
x'&=x\left(-3+\delta+3x\right),\label{eq13a}\\
\delta'&=\delta\left(-\delta+\frac{3}{2}x\right).\label{eq13b}
\end{align}
The phase portrait for the dynamical system
(\ref{eq13a})-(\ref{eq13b}) is presented in Figure~\ref{fig:2}. On
this phase portrait the deS$_{+}$ universe is a global attractor
for expanding universes. On another hand the critical point (3) is a
global repeller representing the Einstein-de Sitter universe. The
saddle point is representing the static Einstein universe.

For the analysis of the behavior of trajectories at infinity we
use the following sets of two projective coordinates:
${\tilde{x}}=\frac{1}{x}$, ${\tilde{\delta}}=\frac{\delta}{x}$ and
${\tilde{X}}=\frac{x}{\delta}$,
${\tilde{\Delta}}=\frac{1}{\delta}$.

\begin{figure}
    \centering
    \includegraphics[width=0.7\linewidth]{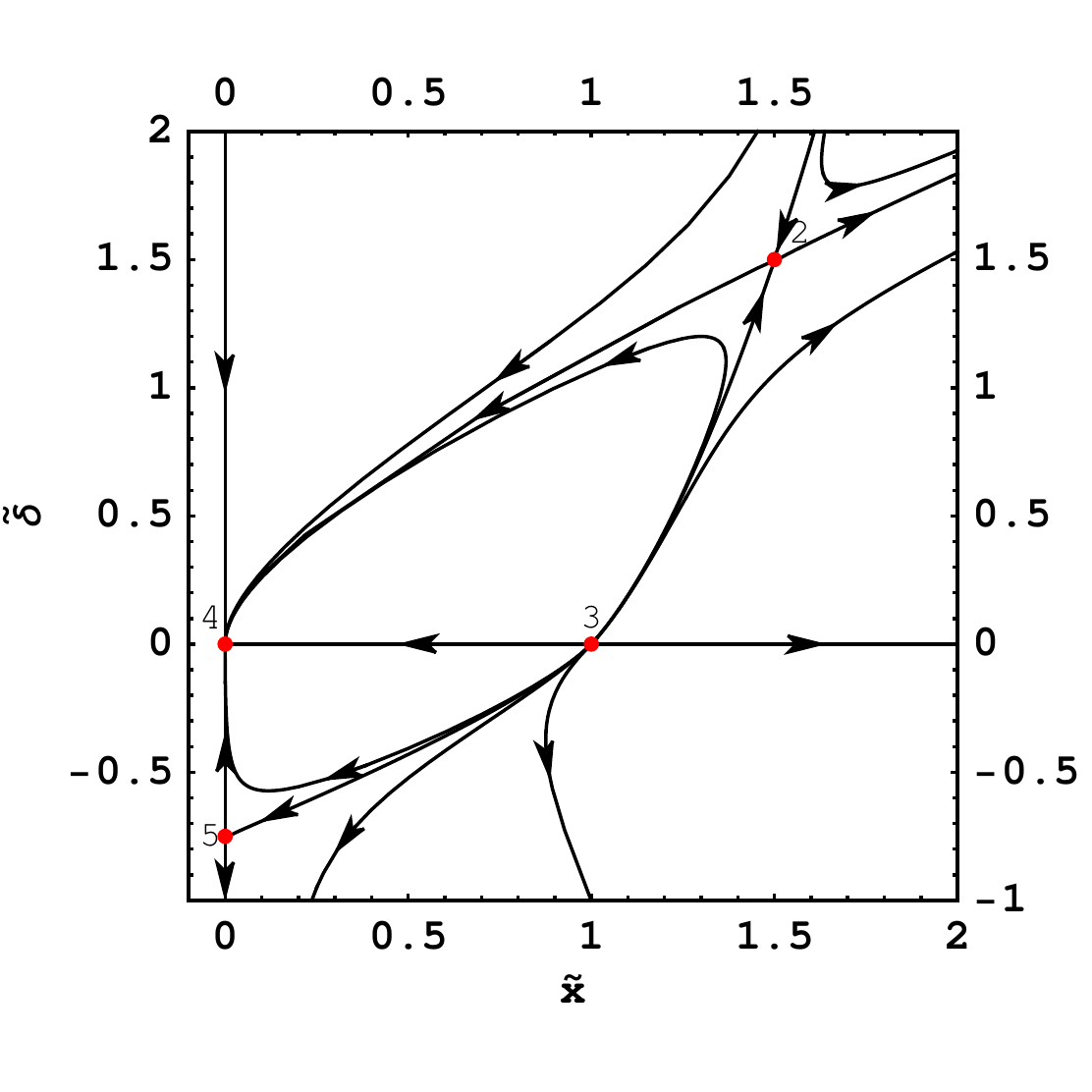}
    \caption{A phase portrait for dynamical system (\ref{eq14a})-(\ref{eq14b}). Critical point (4) (${\tilde{x}}=0$, ${\tilde{\delta}}=0$) and (5) (${\tilde{x}}=0$, ${\tilde{\delta}}=-3/4$) and represents the static universe. Critical point (2) (${\tilde{x}}=3/2$, ${\tilde{\delta}}=3/2$) is a saddle and represents the scaling universe. Critical point (3) (${\tilde{x}}=1$, ${\tilde{\delta}}=0$) is an unstable node and represents the Einstein-de Sitter universe.}
    \label{fig:3}
\end{figure}

The dynamical system in variables ${\tilde{x}}$ and
${\tilde{\delta}}$ covers the behavior of trajectories at infinity
\begin{align}
{\tilde{x}}'&={\tilde{x}}(3{\tilde{x}}-{\tilde{\delta}}-3),\label{eq14a}\\
{\tilde{\delta}}'&={\tilde{\delta}}\left(3{\tilde{x}}-2{\tilde{\delta}}-\frac{3}{2}\right),\label{eq14b}
\end{align}
where $'\equiv {\tilde{x}}\frac{d}{d\tau}$. The phase portrait for
the dynamical system  (\ref{eq14a})-(\ref{eq14b}) is presented in
Figure~\ref{fig:3}.

\begin{figure}
    \centering
    \includegraphics[width=0.7\linewidth]{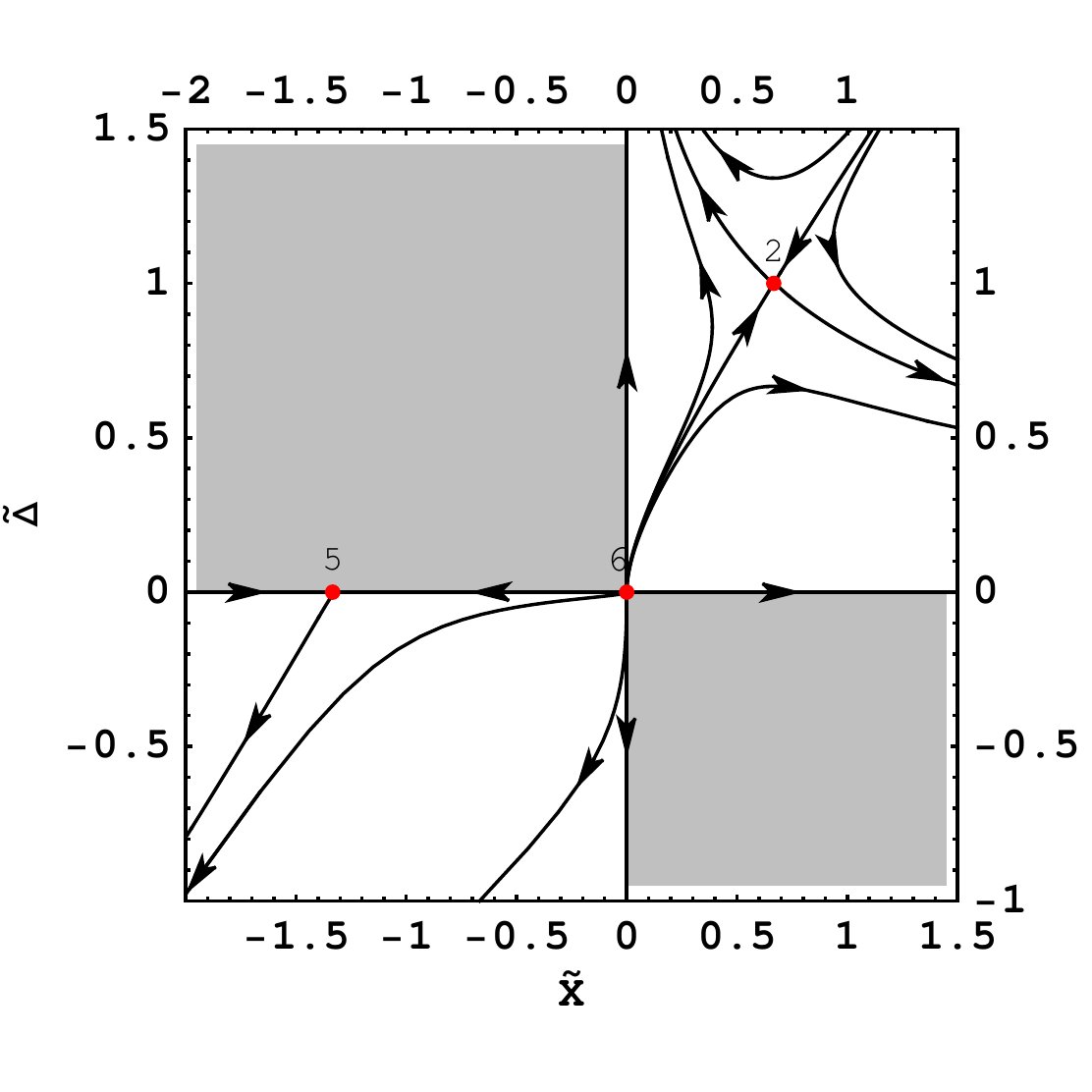}
    \caption{A phase portrait for dynamical system (\ref{eq15a})-(\ref{eq15b}). Critical point (5) (${\tilde{X}}=-4/3$, ${\tilde{\Delta}}=0$) represents the static universe. Critical point (2) (${\tilde{X}}=2/3$, ${\tilde{\Delta}}=1$) is a saddle and represents the scaling universe. Critical point (6) (${\tilde{X}}=0$, ${\tilde{\Delta}}=0$) is an unstable node and represents the de Sitter universe. Note that if ${\tilde{\Delta}}<0$ the arrow of time indicates how the scale factor is decreasing during the evolution.}
    \label{fig:4}
\end{figure}

The dynamical system for variables ${\tilde{X}}$ and
${\tilde{\Delta}}$ is described by the following equations
\begin{align}
{\tilde{X}}'&={\tilde{X}}\left(-3{\tilde{\Delta}}+\frac{3}{2}{\tilde{X}}+2\right),\label{eq15a}\\
{\tilde{\Delta}}'&={\tilde{\Delta}}\left(1-\frac{3}{2}{\tilde{X}}\right),\label{eq15b}
\end{align}
where $'\equiv {\tilde{\Delta}}\frac{d}{d\tau}$. The phase
portrait for the dynamical system (\ref{eq15a})-(\ref{eq15b}) is
presented in Figure~\ref{fig:4}.

\begin{figure}
    \centering
    \includegraphics[width=0.7\linewidth]{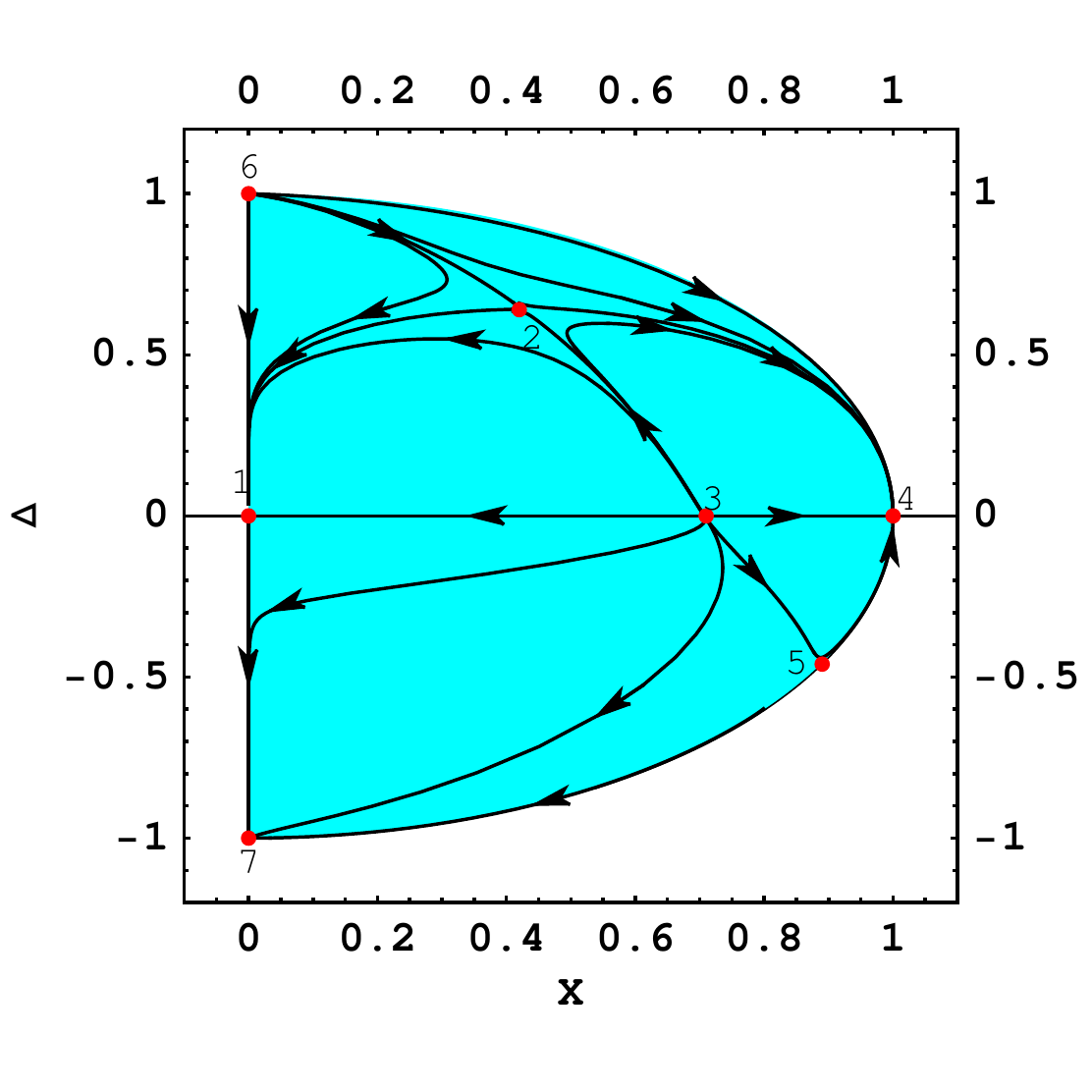}
    \caption{A phase portrait for dynamical system (\ref{eq19a})-(\ref{eq19b}). Critical point (1) represents the de Sitter universe. Critical point (2) is a saddle and represents the scaling universe. Critical point (3) is an unstable node and represents the Einstein-de Sitter universe. Critical point (4) represents the static universe. Critical point (5) represents the static universe. Critical point (6) is an unstable node and represents the de Sitter universe. Note that if ${\Delta}<0$ the arrow of time indicates how the scale factor is decreasing during the evolution.}
    \label{fig:8}
\end{figure}

We use also the Poincar{\'e} sphere to analyze critical points in
the infinity. We define variables
\begin{equation}
X=\frac{x}{\sqrt{1+x^2+\delta^2}}, \quad
\Delta=\frac{\delta}{\sqrt{1+\delta^2+x^2}}
\end{equation}
and in these variables the dynamical system has the following form
\begin{equation}
X'=X\left[-\Delta^2(\frac{3}{2}X-\Delta)+(1-X^2)
(3X+\Delta-3\sqrt{1-X^2-\Delta^2})\right],\label{eq19a}
\end{equation}
\begin{equation}
\Delta'=\Delta\left[(1-\Delta^2)(\frac{3}{2}X-\Delta)-X^2
(3X+\Delta-3\sqrt{1-X^2-\Delta^2})\right],\label{eq19b}
\end{equation}
where $'\equiv \sqrt{1-X^2-\Delta^2}\frac{d}{d\tau}$. The phase
portrait for the dynamical system (\ref{eq19a})-(\ref{eq19b}) is
presented in Figure~\ref{fig:8}. Critical points for autonomous
dynamical systems (\ref{eq13a})-(\ref{eq13b}),
(\ref{eq14a})-(\ref{eq14b}), (\ref{eq15a})-(\ref{eq15b}) are
completed in Table~\ref{table:4}.

\begin{table}
    \caption{Critical points for autonomous dynamical systems (\ref{eq13a})-(\ref{eq13b}), (\ref{eq14a})-(\ref{eq14b}), (\ref{eq15a})-(\ref{eq15b}), their type and cosmological interpretation.}
    \label{table:4}
    \begin{center}
        \begin{tabular}{llll} \hline
            No. & critical point & type of critical point & type of universe \\ \hline \hline
            1 & $x=0$, $\delta=0$ & saddle-node & de Sitter universe without diffusion effect \\
            2 & $x=2/3$, $\delta=1$ & saddle & scaling universe \\
              & ($\tilde{x}=3/2$, $\tilde{u}=3/2$) & & \\
              & ($\tilde{X}=2/3$, $\tilde{U}=1$) & & \\  
            3 & $x=1$, $\delta=0$ & unstable node & Einstein-de Sitter universe without diffusion effect\\
              & ($\tilde{x}=1$, $\tilde{u}=0$) & & \\
            4 & $\tilde{x}=0$, $\tilde{\delta}=0$ & stable node & static universe \\
            5 & $\tilde{X}=0$, $\tilde{\Delta}=-3/4$ & saddle & static universe\\
            6 & $\tilde{X}=0$, $\tilde{\Delta}=0$ & unstable node & de Sitter universe with diffusion effect \\
            \\ \hline
        \end{tabular}
    \end{center}
\end{table}

\subsection{Dynamics of the model at the late time ($ma \to \infty$ )}

As can be seen from eq.~(\ref{eq:29}) the relativistic model of the dark
matter consists of two fluids first with $\tilde{w}=0$
and the second with  $\tilde{w}=\frac{2}{3}$. So, in the approximation
$ma \to \infty$ and $w=-1$ we have according to eqs.~(\ref{eq:28}), (\ref{eq:32}) and (\ref{eq:36}) the following DM and DE
continuity equations
\begin{gather}
\dot{\rho}_{\text{dm}}+5\rho_{\text{dm}} H=\gamma a^{-3}+2Zm Ha^{-3},\\
\dot{\rho}_{\text{de}}=-\gamma a^{-3}.
\end{gather}
We define the  variables
\begin{equation}
x=\frac{\rho_{\text{dm}}}{3H^2},\quad
y=\frac{\rho_{\text{de}}}{3H^2},\quad u=\frac{(2Zm
)a^{-3}}{\rho_{\text{dm}}}\quad \text{and} \quad
\delta=\frac{\gamma a^{-3}}{H\rho_{\text{dm}}}.\label{eq21}
\end{equation}
If we use the  variables (\ref{eq21}) and time $\tau=\ln a$ then
we obtain the following dynamical system
\begin{align}
x'&=x(-5+\delta+u-2 \frac{\dot{H}}{H^2}),\label{eq22a}\\
y'&=-x(\delta+u)-2y \frac{\dot{H}}{H^2},\\
u'&=u(2-\delta-u),\\
\delta'&=\delta(2-\delta-u-\frac{\dot{H}}{H^2}),\label{eq22d}
\end{align}
where $ '\equiv \frac{d}{d\tau}$ and $\frac{\dot{H}}{H^2}=-\frac{1}{2}x(5-u)$.
Because $x+y=1$ ($\Omega_{\text{dm}}+\Omega_{\text{de}}=1$) then
the dynamical system (\ref{eq22a})-(\ref{eq22d}) reduces to the
three-dimensional dynamical system.

\begin{figure}
    \centering
    \includegraphics[width=0.7\linewidth]{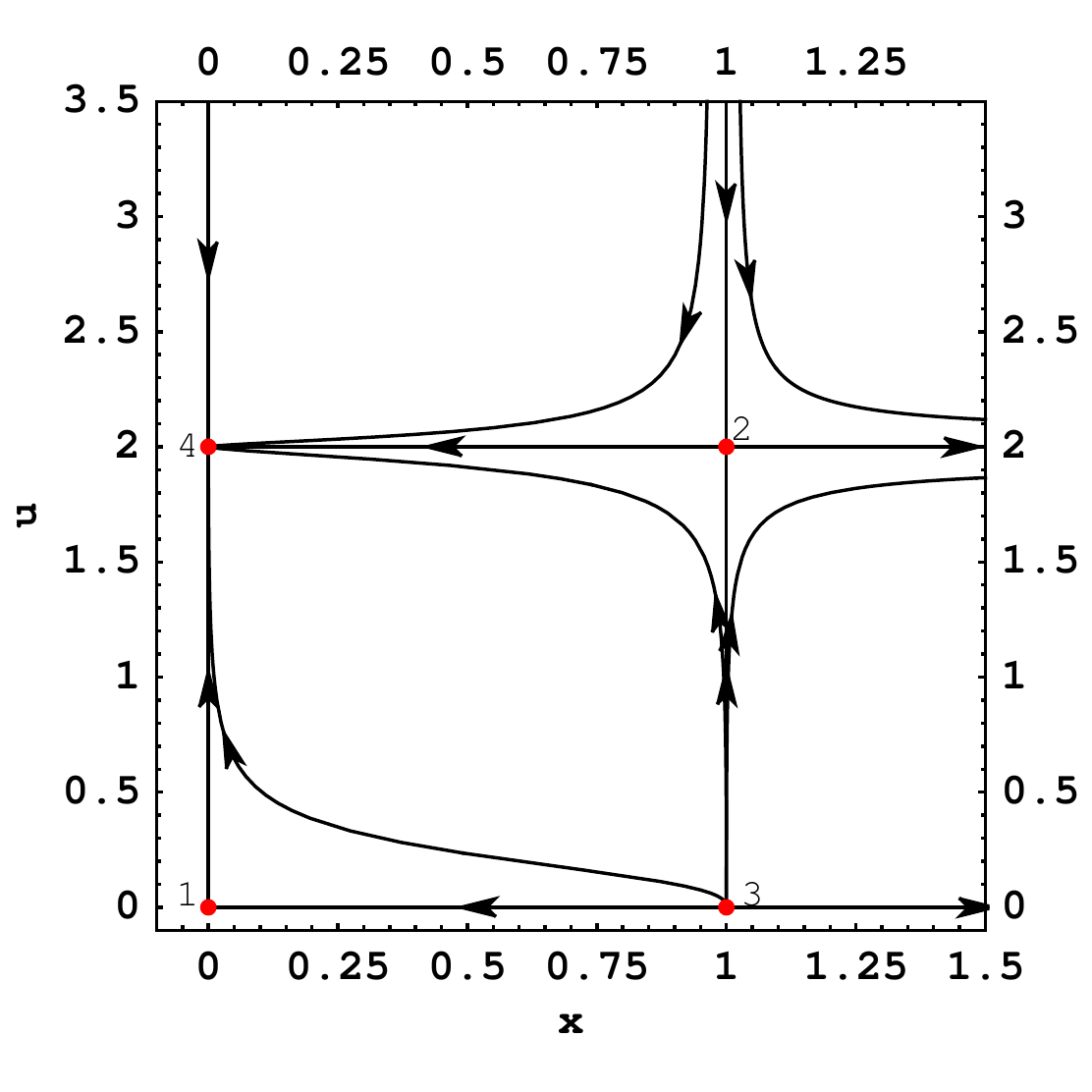}
    \caption{A phase portrait for dynamical system (\ref{eq16a})-(\ref{eq16b}). Critical point (1) ($x=0$, $u=0$) represents the de Sitter universe without the diffusion effect. Critical point (2) ($x=1$, $u=2$) is a saddle type and represents the scaling universe. Critical point (3) ($x=1$, $u=0$) is an unstable node and represents the Einstein-de Sitter universe without the diffusion effect. The critical point (4) is representing the Einstein-de Sitter without the diffusion effect.}
    \label{fig:5}
\end{figure}

The dynamical system (\ref{eq22a})-(\ref{eq22d}) has the invariant
submanifold $\{\frac{\dot{H}}{H^2}=0\}$, which is the set
$\{x\colon x=0\}$ or $\{u\colon u=5\}$. There is also an interesting submanifold
$\delta=0$. On the invariant submanifold $\delta=0$ the dynamical
system (\ref{eq22a})-(\ref{eq22d}) reduces to
\begin{align}
x'&=x(u+5(x-1)-xu),\label{eq16a}\\
u'&=u(2-u).\label{eq16b}
\end{align}
The phase portrait for the dynamical system
(\ref{eq16a})-(\ref{eq16b}) is presented in Figure~\ref{fig:5}.
Note that critical point (1) is representing the deS$_{+}$
universe without the diffusion effect. On the other hand the de
Sitter universe without diffusion is represented by saddle
critical point. Therefore the model with diffusion is generic in
the class of all trajectories.

For the analysis the behavior of trajectories at infinity we use
the following two sets of projective coordinates:
${\tilde{x}}=\frac{1}{x}$, ${\tilde{u}}=\frac{u}{x}$ and
${\tilde{X}}=\frac{x}{u}$, ${\tilde{U}}=\frac{1}{u}$ .

\begin{figure}
    \centering
    \includegraphics[width=0.7\linewidth]{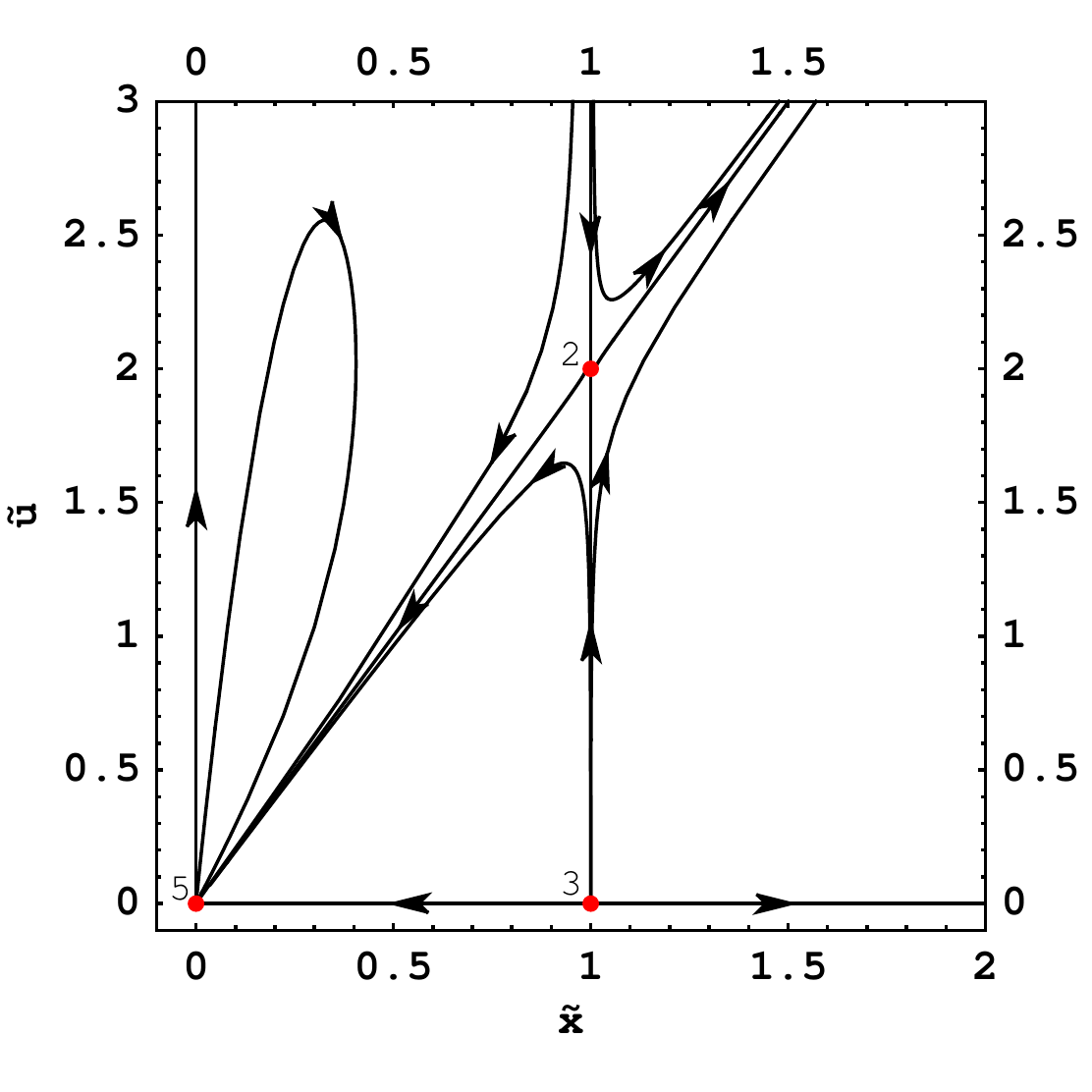}
    \caption{A phase portrait for dynamical system (\ref{eq17a})-(\ref{eq17b}). Critical point (5) (${\tilde{x}}=0$, ${\tilde{u}}=0$) represents the static universe. Critical point (2) (${\tilde{x}}=1$, ${\tilde{u}}=2$) is a saddle and represents the scaling universe. Critical point (3) (${\tilde{x}}=1$, ${\tilde{u}}=0$) is an unstable node and represents the Einstein-de Sitter universe. Note that the Einstein-de Sitter universe is fully dominated by dark matter. It is an attractor solution as well as the de Sitter which one can see in Figure~\ref{fig:5}}
    \label{fig:6}
\end{figure}

The dynamical system for variables ${\tilde{x}}$ and ${\tilde{u}}$
is expressed by
\begin{align}
{\tilde{x}}'&={\tilde{x}}(5\tilde{x}({\tilde{x}}-1)+{\tilde{u}}(1-\tilde{x})),\label{eq17a}\\
{\tilde{u}}'&={\tilde{u}}\left({\tilde{x}}(7\tilde{x}-5)+\tilde{u}(1-2\tilde{x})+\tilde{u}\right),\label{eq17b}
\end{align}
where $'\equiv {\tilde{x}}^2\frac{d}{d\tau}$. The phase portrait
for above dynamical system is presented in Figure~\ref{fig:6}. In
comparison to the phase portrait in Figure~\ref{fig:5} a new
critical point (5) is emerging. It is representing the Einstein-de
Sitter universe fully dominated by dark matter.

\begin{figure}
    \centering
    \includegraphics[width=0.7\linewidth]{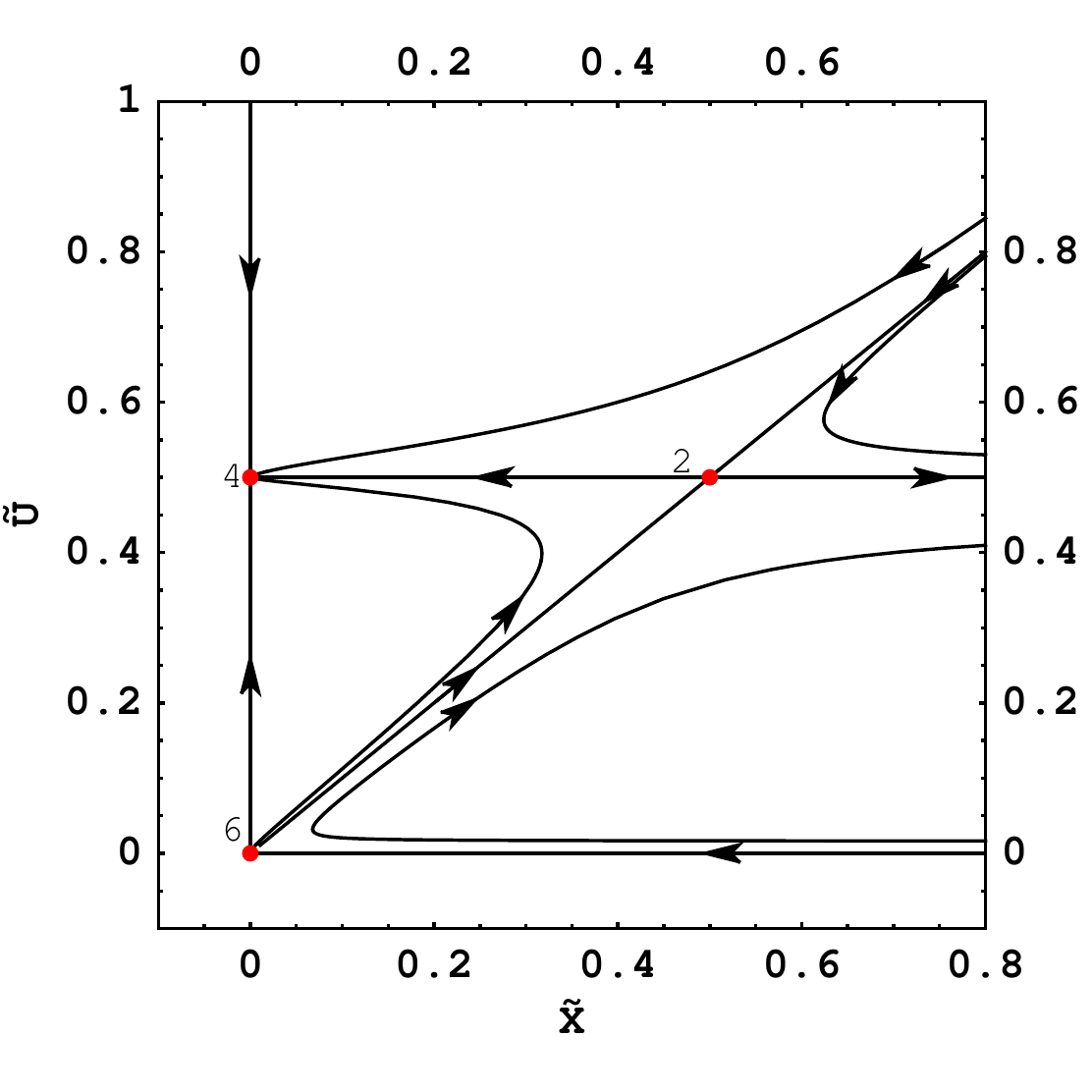}
    \caption{A phase portrait for dynamical system (\ref{eq18a})-(\ref{eq18b}). Critical point (2) (${\tilde{X}}=1/2$, ${\tilde{U}}=1/2$) is a saddle and represents the scaling universe. Critical point (6) (${\tilde{X}}=0$, ${\tilde{U}}=0$) is an unstable node and represents the de Sitter universe. At this critical point the effect of diffusion are important. On the other hand the critical point (6) is an unstable stationary solution in which the effect of the non-zero term ($Zm$) vanishes.}
    \label{fig:7}
\end{figure}

The dynamical system for variables ${\tilde{X}}$ and
${\tilde{\Delta}}$ is described by the following equations
\begin{align}
{\tilde{X}}'&={\tilde{X}}\left({\tilde{U}}(2-7\tilde{U})+{\tilde{X}}(5\tilde{U}-1)\right),\label{eq18a}\\
{\tilde{U}}'&={\tilde{U}}^2\left(1-2{\tilde{U}}\right),\label{eq18b}
\end{align}
where $'\equiv {\tilde{U}}^2\frac{d}{d\tau}$. The phase portrait
for the dynamical system (\ref{eq18a})-(\ref{eq18b}) is presented
in Figure~\ref{fig:7}. Note  the de Sitter universe represented by
critical point (1) which is a stationary universe without effect of
diffusion is a global attractor.

Critical points for autonomous dynamical systems
(\ref{eq16a})-(\ref{eq16b}), (\ref{eq17a})-(\ref{eq17b}),
(\ref{eq18a})-(\ref{eq18b}) are completed in Table~\ref{table:3}.

\begin{table}
    \caption{Critical points for autonomous dynamical systems (\ref{eq16a})-(\ref{eq16b}), (\ref{eq17a})-(\ref{eq17b}), (\ref{eq18a})-(\ref{eq18b}), their types and cosmological interpretations.}
    \label{table:3}
    \begin{center}
        \begin{tabular}{llll} \hline
            No & critical point & type of critical point & type of universe \\ \hline \hline
            1 & $x=0$, $u=0$ & saddle & de Sitter universe without diffusion effect \\
            2 & $x=1$, $u=2$ & saddle & scaling universe \\
	          & ($\tilde{x}=1$, $\tilde{u}=2$) & & \\
	          & ($\tilde{X}=1/2$, $\tilde{U}=1/2$) & & \\  
            3 & $x=1$, $u=0$ & unstable node & Einstein-de Sitter universe without diffusion effect\\
              & ($\tilde{x}=1$, $\tilde{u}=0$) & & \\
            4 & $x=0$, $u=2$ & stable node & de Sitter universe without diffusion effect \\
              & ($\tilde{X}=0$, $\tilde{U}=1/2$) & & \\
            5 & $\tilde{x}=0$, $\tilde{u}=0$ & stable node & static universe\\
            6 & $\tilde{X}=0$, $\tilde{U}=0$ & unstable node & de Sitter universe without diffusion effect \\
             \\ \hline
        \end{tabular}
    \end{center}
\end{table}



\begin{figure}
    \centering
    \includegraphics[width=0.7\linewidth]{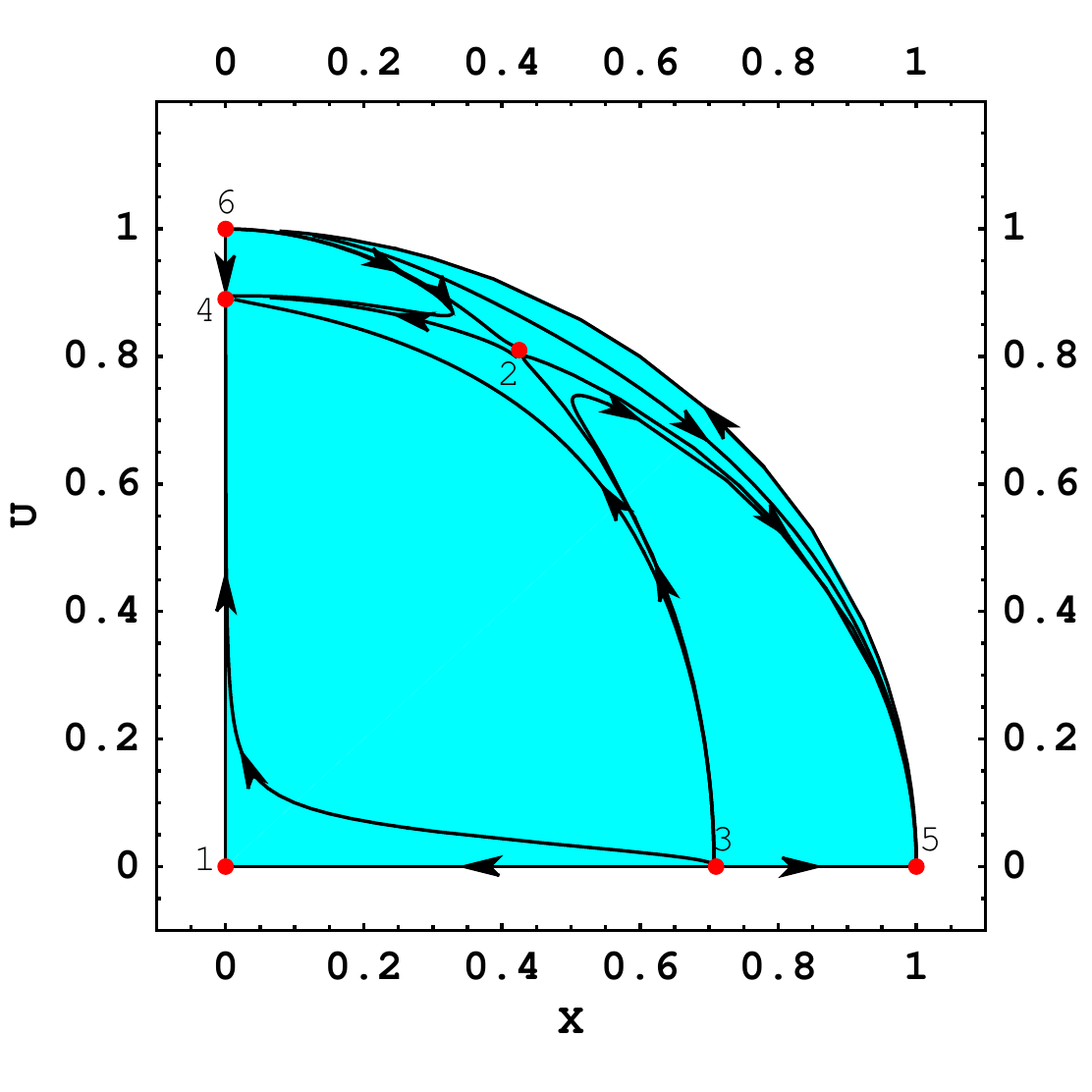}
    \caption{A phase portrait for dynamical system (\ref{eq20a})-(\ref{eq20b}). Critical point (1) represents the de Sitter universe without the diffusion effect. Critical point (2) is a saddle type and represents the scaling universe. Critical point (3) is an unstable node and represents the Einstein-de Sitter universe without the diffusion effect. The critical point (4) is representing the Einstein-de Sitter with the diffusion effect. Critical point (5) represents the static universe. Critical point (6) is an unstable node and represents the de Sitter universe.}
    \label{fig:9}
\end{figure}

We apply also the Poincar{\'e} sphere to this system in order to
analyze critical points at infinity. We define the  variables
\begin{equation}
    X=\frac{x}{\sqrt{1+x^2+u^2}}, \quad U=\frac{u}{\sqrt{1+x^2+u^2}}
\end{equation}
In these variables the dynamical system has the following form
\begin{align}
    X'&=X\left[U^2 \sqrt{1-X^2-U^2}(U-2\sqrt{1-X^2-U^2}) \right. \nonumber \\
    &+ \left. (1-X^2) (\sqrt{1-X^2-U^2}(5X+U)-5(1-X^2-U^2)-XU)\right],\label{eq20a}\\
    U'&=U\left[(1-U^2)\sqrt{1-X^2-U^2}(2\sqrt{1-X^2-U^2}-U) \right. \nonumber\\
    &- \left. X^2 (\sqrt{1-X^2-U^2}(5X+U)-5(1-X^2-U^2)-XU)\right],\label{eq20b}
\end{align}
where $'\equiv (1-X^2-U^2)\frac{d}{d\tau}$. The phase
portrait for the dynamical system (\ref{eq20a})-(\ref{eq20b}) is
presented in Figure~\ref{fig:9}.

\section{Statistical analysis}

\subsection{Introduction}

In this section we use astronomical observations for low redshifts
such as the SNIa, BAO, measurements of $H(z)$ for galaxies and the
Alcock-Paczy{\'n}ski test. We do not use the observation for high redshifts such as CMB.
\begin{equation}
\ln L_{\text{SNIa}} = -\frac{1}{2} [A - B^2/C + \ln(C/(2 \pi))],
\end{equation}
where $A=
(\mathbf{\mu}^{\text{obs}}-\mathbf{\mu}^{\text{th}})\mathbb{C}^{-1}(\mathbf{\mu}^{\text{obs}}-\mathbf{\mu}^{\text{th}})$,
$B=
\mathbb{C}^{-1}(\mathbf{\mu}^{\text{obs}}-\mathbf{\mu}^{\text{th}})$,
$C=\text{Tr} \mathbb{C}^{-1}$ and $\mathbb{C}$ is a covariance
matrix for SNIa. The distance modulus is expressed by
$\mu^{\text{obs}}=m-M$ (where $m$ is the apparent magnitude and
$M$ is the absolute magnitude of SNIa) and $\mu^{\text{th}} = 5
\log_{10} D_L +25$ (where the luminosity distance is $D_L= c(1+z)
\int_{0}^{z} \frac{d z'}{H(z)}$).

We also use BAO observations such as Sloan Digital Sky Survey
Release 7 (SDSS DR7) dataset at $z = 0.275$
\cite{Percival:2009xn}, 6dF Galaxy Redshift Survey measurements at
redshift $z = 0.1$ \cite{Beutler:2011hx}, and WiggleZ
measurements at redshift $z = 0.44, 0.60, 0.73$
\cite{Blake:2012pj}. The likelihood function is expressed by the
formula
\begin{equation}
\ln L_{\text{BAO}} = -
\frac{1}{2}\left(\mathbf{d}^{\text{obs}}-\frac{r_s(z_d)}{D_V(\mathbf{z})}\right)\mathbb{C}^{-1}\left(\mathbf{d}^{\text{obs}}-\frac{r_s(z_d)}{D_V(\mathbf{z})}\right),
\end{equation}
where $r_s(z_d)$ is the sound horizon at the drag epoch
\cite{Eisenstein:1997ik}.

For the Alcock-Paczynski test
\cite{Alcock:1979mp,Lopez-Corredoira:2013lca} we use the
likelihood function
\begin{equation}
\ln L_{AP} =  - \frac{1}{2} \sum_i \frac{\left(
AP^{th}(z_i)-AP^{obs}(z_i) \right)^2}{\sigma^2}.
\end{equation}
where $AP(z)^{\text{th}} \equiv \frac{H(z)}{z} \int_{0}^{z}
\frac{dz'}{H(z')}$ and $AP(z_i)^{\text{obs}}$ are observational
data
\cite{Sutter:2012tf,Blake:2011ep,Ross:2006me,Marinoni:2010yoa,daAngela:2005gk,Outram:2003ew,Anderson:2012sa,Paris:2012iw,Schneider:2010hm}.

In addition,  we are applying measurements of the Hubble parameter
$H(z)$ of galaxies from
\cite{Simon:2004tf,Stern:2009ep,Moresco:2012jh}. In this case the
likelihood function is expressed by
\begin{equation}\label{hz}
\ln L_{H(z)} = -\frac{1}{2} \sum_{i=1}^{N}  \left
(\frac{H(z_i)^{\text{obs}}-H(z_i)^{\text{th}}}{\sigma_i
}\right)^2.
\end{equation}

The final likelihood function is in the following form
\begin{equation}
L_{\text{tot}} = L_{\text{SNIa}} L_{\text{BAO}} L_{\text{AP}}
L_{H(z)}.
\end{equation}

We use our own code CosmoDarkBox to estimate the model parameters.
This code applies the Metropolis-Hastings algorithm
\cite{Metropolis:1953am,Hastings:1970aa} and the dynamical system
formulation of model dynamics to obtain the likelihood function
\cite{Hu:1995en,Eisenstein:1997ik}. The dynamical system
formulation of the cosmological dynamics developed in sec.~5 plays
a crucial role in our method of estimation. We solve the system
numerically using the Monte Carlo method and than put this
solution to the corresponding expression for observables in our
model.

For comparison models with diffusion with the $\Lambda$CDM model, we use Bayesian information criterion (BIC) \cite{Schwarz:1978ed,Kass:1995bf}. The BIC is defined as
\begin{equation}
\text{BIC}=-2\ln L+j \ln n,
\end{equation}
where $L$ is the maximum of the likelihood function, $j$ is the number of model parameters (in this paper for our models $j=3$ and for $\Lambda$CDM $j=2$) and $n$ is number of data points (in this paper $n=622$).

\subsection{Model of DM-DE interaction and $\tilde{w}=0$}

Let us consider the model of DM-DE interaction and with dark
matter in the form of dust. We present a statistical analysis of
the model parameters such as $H_0$,
$\Omega_{\text{dm},0}=\frac{\rho_{\text{dm},0}}{3H_0^2}$, where $\rho_{\text{dm},0}$ is the present value of dark matter and
$\Omega_{\gamma,0}=\frac{\gamma}{3H_0^2}\int^T dt$, where $T$
is the present age of the Universe. We must have
$\Omega_{\gamma}\geq 0$ because $\gamma\geq 0$ for a diffusion.

The Friedmann equation for $\tilde{w}=0$ in terms of the present values of the density parameters takes the
form
\begin{equation}
\frac{H^2}{H_0^2}=\Omega_{\text{cm,0}}a^{-3}+\frac{\Omega_{\gamma,0}}{\int^T dt}a^{-3}\int^t
dt+\Omega_{\text{b,0}}a^{-3}+\Omega_{\text{de}}(0)-\frac{\bar\Omega_{\gamma,0}}{\int^T
a^{-3} dt}\int^t a^{-3} dt,
\end{equation}
where $\Omega_{\text{cm},0}=\frac{\rho_{\text{cm},0}}{3H_0^2}$, where $\rho_{\text{cm},0}$ is the present value of the conservative part of dark matter, which scales as $a^{-3}$, $\bar\Omega_{\gamma,0}=\frac{\gamma}{3H_0^2}\int^T a^{-3} \, dt$.

In these estimation we use formulation of dynamics in the form of
a two-dimensional non-autonomous system with the redshift variable
$z$. This model possesses three parameters $\gamma$, $H_0$ and
$\Omega_{\text{m}}(z=0)=\Omega_{\text{m},0}$
\begin{equation}
\Omega_{\text{m}}'=\frac{3}{1+z}\Omega_{\text{m}}-\frac{\gamma}{3H_0^3}P
(1+z)^{2},
\end{equation}
\begin{equation}
P'=-\frac{3}{2}\frac{1}{1+z}\Omega_{\text{m},0}P^3,
\end{equation}
where $P=\frac{H_0}{H}$ and $'\equiv z$.

\begin{figure}[ht]
   \centering
   \includegraphics[scale=1]{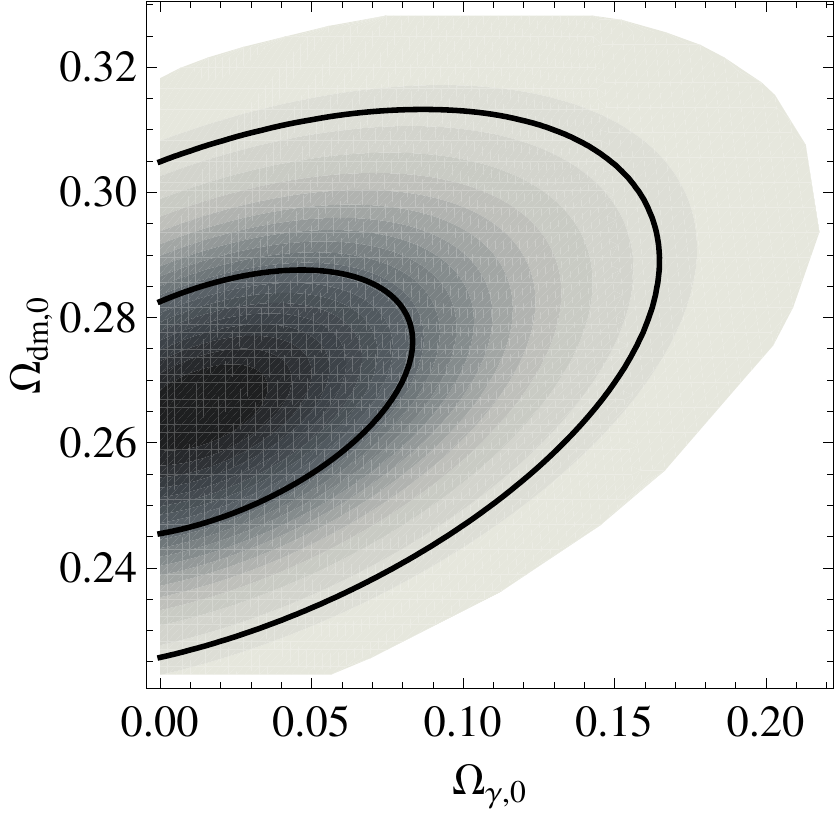}
   \caption{The intersection of the likelihood function of two model parameters ($\Omega_{\text{dm},0}$, $\Omega_{\gamma,0}$), for the case of the model of DM-DE interaction and $\tilde{w}=0$, with the marked $68\%$ and $95\%$ confidence levels for SNIa+BAO+$H(z)$+AP test. $\Omega_{\text{dm},0}$ is the present value of dark matter.}
  \label{fig:1}
\end{figure}

Statistical results are presented in Table~\ref{table:1}.
Figure~\ref{fig:1} shows the likelihood function with $68\%$ and
$95\%$ confidence level projections on the plane
($\Omega_{\text{dm},0}$, $\Omega_{\gamma}$). For this case the
value of reduced $\chi^2$ is equal 0.187767.

The value of BIC, for this model is equal BIC$_1$=135.527. Because BIC for the $\Lambda$CDM model is equal BIC$_{\Lambda\text{CDM}}$=129.105, $\Delta\text{BIC}=\text{BIC}_1- \text{BIC}_{\Lambda\text{CDM}}$ is equal 6.421. If that a value of $\Delta$BIC is more than 6, the evidence for the model is strong \cite{Kass:1995bf}. Consequently, the evidence in favor of the $\Lambda$CDM model is strong in comparison to our model. 

\begin{table}
    \caption{The best fit and errors for the estimated model for SNIa+BAO+$H(z)$+AP test with $H_0$ from the interval (65.0 (km/(s Mpc)), 71.0 (km/(s Mpc))), $\Omega_{\text{dm},0}$ from the interval $(0.25, 0.40)$, $\Omega_{\gamma,0}$ from the interval $(0.00, 0.20)$ $\Omega_{\text{b},0}$ is assumed as 0.048468. The value of reduced $\chi^2$ is equal 0.187767.}
    \label{table:1}
    \begin{center}
        \begin{tabular}{llll} \hline
            parameter & best fit & $ 68\% $ CL & $ 95\% $ CL  \\ \hline \hline
            $H_0$ & 67.97 km/(s Mpc) & $\begin{array}{c}
            +0.75 \\ -0.72
            \end{array}$ & $\begin{array}{c}
            +1.57 \\ -1.45
            \end{array}$ \\ \hline
            $\Omega_{\text{dm},0}$ & 0.2658 & $\begin{array}{c}
            +0.0223 \\ -0.0208
            \end{array}$ & $\begin{array}{c}
            +0.0485 \\ -0.0415
            \end{array}$ \\ \hline
            $\Omega_{\gamma,0}$ & 0.0135 & $\begin{array}{c}
            +0.0735 \\ -0.0135
            \end{array}$ & $\begin{array}{c}
            +0.1570 \\ -0.0135
            \end{array}$ \\ \hline
        \end{tabular}
    \end{center}
\end{table}

\subsection{Model with DM-DE interaction for $m a \to \infty$}
Let us consider a late time behavior of the universe. For the case
$ma \to \infty$ we estimated values of cosmological parameters
such as $\Omega_{\gamma,0}=\frac{\gamma}{3H_0^2}\int^T a^2 dt$, $\Omega_{Zm,0}=\frac{Zm}{3H_0^2}$, $H_0$
and $\gamma$. The formula for the Friedmann equation in terms
of the present values of the density parameters is in the form
\begin{equation}
\frac{H^2}{H_0^2}=\Omega_{\text{dm},0}a^{-5}+\frac{\Omega_{\gamma,0}}{\int^T
a^2 dt}a^{-5}\int^t a^2 dt+\Omega_{Zm}
a^{-3}+\Omega_{\text{de}}(0)-\frac{\bar\Omega_{\gamma,0}}{\int^T
a^{-3} dt}\int^t a^{-3} dt,
\end{equation}
where $\bar\Omega_{\gamma,0}=\frac{\gamma}{3H_0^2}\int^T a^{-3}
dt$ and $\Omega_{\text{dm},0}$ is the present value of the part of dark matter,
which scales as $a^{-5}$.

\begin{figure}[ht]
    \centering
    \includegraphics[scale=1]{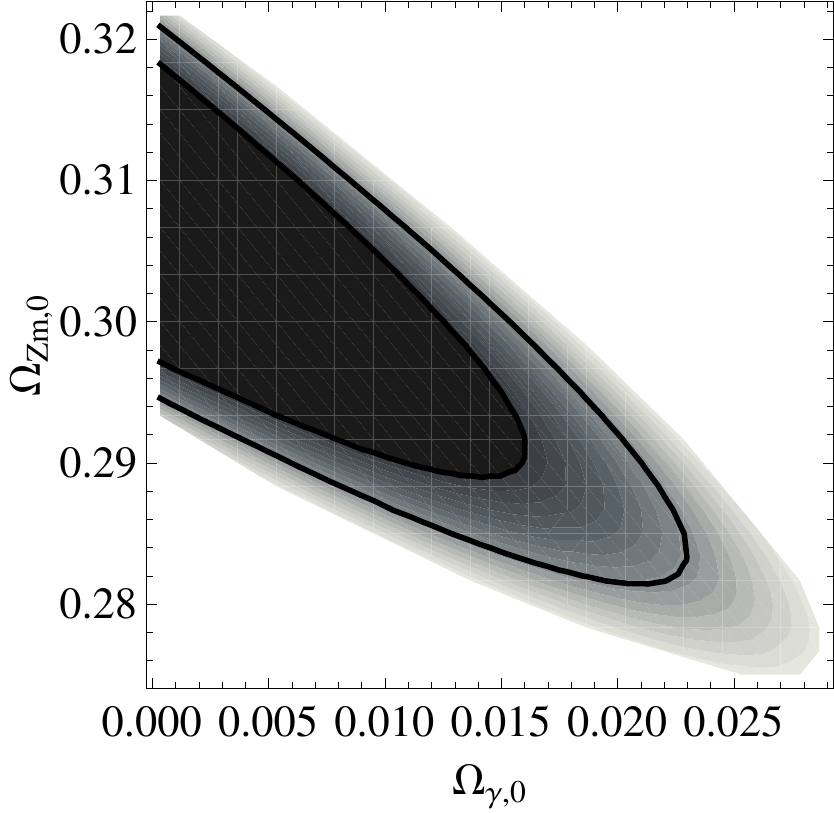}
    \caption{The intersection of the likelihood function of two model parameters ($\Omega_{\text{Zm},0}$, $\Omega_{\gamma,0}$), for the case of the model with DM-DE interaction for $ma \to \infty$, with the marked $68\%$ and $95\%$ confidence levels for SNIa+BAO+$H(z)$+AP test.}
    \label{fig:10}
\end{figure}

The results of our analysis of the model are completed in Table
\ref{table:2}. Figure~\ref{fig:10} shows the likelihood function
with the $68\%$ and $95\%$ confidence level projections on the
plane ($\Omega_{\text{dm},0}$, $\Omega_{\gamma}$). For this case
the value of reduced $\chi^2$ is equal 0.188201.

The value of BIC, for this model is equal BIC$_2$=135.795 Because BIC for the $\Lambda$CDM model is equal BIC$_{\Lambda\text{CDM}}=129.105$, $\Delta\text{BIC}=\text{BIC}_2-\text{BIC}_{\Lambda\text{CDM}}$ is equal 6.690. If that a value of $\Delta$BIC is more than 6, the evidence for the model is strong \cite{Kass:1995bf}. Consequently, the evidence in favor of the $\Lambda$CDM model is strong in comparison to our model. 
\begin{table}
    \caption{The best fit and errors for the estimated model with $w=2/3$ for SNIa+BAO+$H(z)$+AP test with $\Omega_{Zm,0}$ from the interval $(0.22, 0.38)$, $\Omega_{\gamma,0}$ from the interval $(0.0, 0.03)$, $\gamma$ from the interval $(0.00 \text{(100 km/(s Mpc))}^3, 0.500 \text{(100 km/(s Mpc))}^3)$ and $H_0$ from the interval (65.0 (km/(s Mpc)), 71.0 (km/(s Mpc))). $\Omega_{\text{b},0}$ is assumed as 0.048468. The value of reduced $\chi^2$ is equal 0.188201.}
    \label{table:2}
    \begin{center}
        \begin{tabular}{llll} \hline
            parameter & best fit & $ 68\% $ CL & $ 95\% $ CL  \\ \hline \hline
            $H_0$ & 68.04 & $\begin{array}{c}
            +0.73 \\ -0.70
            \end{array}$ & $\begin{array}{c}
            +1.27 \\ -1.25
            \end{array}$ \\ \hline
            $\Omega_{\gamma,0}$ & 0.0106 & $\begin{array}{c}
            +0.0082 \\ -0.0106
            \end{array}$ & $\begin{array}{c}
            +0.0137 \\ -0.0106
            \end{array}$ \\ \hline
            $\Omega_{Zm,0}$ & 0.2943 & $\begin{array}{c}
            +0.0356 \\ -0.0077
            \end{array}$ & $\begin{array}{c}
            +0.0536 \\ -0.0231
            \end{array}$ \\ \hline
            $\gamma$ & 0.0299 & $\begin{array}{c}
            +0.2198 \\ -0.0299
            \end{array}$ & $\begin{array}{c}
            +0.4555 \\ -0.0299
            \end{array}$ \\ \hline
        \end{tabular}
    \end{center}
\end{table}

\begin{figure}[ht]
	\centering
	\includegraphics[scale=1]{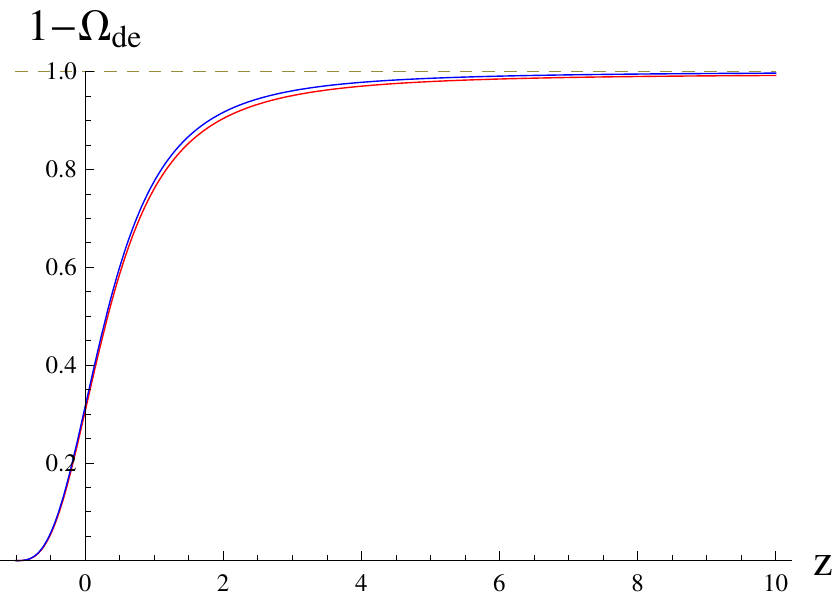}
	\caption{The diagram presents the evolution of $1-\Omega_{de}(z)=\frac{\Omega_{\text{m},0}f(z)}{H^2(z)/H_0^2}$, where $z$ is redshift, $\Omega_{\text{m},0}f(z)=\frac{\rho_{\text{m}}(z)}{3H_0^2}$ and $f(0)=1$, for the first model (blue line) and for the second model (red line). We assume the best fit values of model parameters (see Table \ref{table:1} and \ref{table:2}). Note that, for the early universe, for the both models, $1-\Omega_{de}(z)$ is going to a constant (the horizontal asymptotics equals one). This means, that $\Omega_{de}(z)$ for the high redshifts is negligible.}
	\label{fig:11}
\end{figure}

We can compare the behavior of $\Omega_{\text{de}}$ for our models with others models of the early dark energy. In Doran and Robbers model \cite{Doran:2006kp} the fractional dark energy density is assumed as a constant, which is different from zero, for the early time universe. This means that $\Omega_{de}(z)$ cannot be negligible for the early universe for this model. In our models, $\Omega_{\text{de}}$ approaches zero for the high redshifts (see Figure \ref{fig:11}) and $\Omega_{\text{de}}$ is negligible for the early universe. In consequence, we do not use the high redshift astronomical observations, such as CMB, to fit values of model parameters for our models.

\section{Conclusion}

In this paper we studied the dynamics of DM-DE interaction with
the relativistic diffusion process. For this aim we used the
dynamical system methods, which enable us to study all evolutional
scenarios admissible for all initial conditions. We show that
dynamics of our model reduces to the three-dimensional dynamical system, which in order is
investigated on an invariant two-dimensional submanifold. From our
dynamical analysis the dynamics is free from the difficulties,
which are present in Alho et al.'s models with diffusion
\cite{Alho:2014ola}, namely there is no non-physical trajectories
crossing the boundary set $\rho_{\text{m}}=0$
\cite{Stachowski:2016dfi}.

The model is tested by astronomical data in two cases of dark
matter in the domain of low redshifts (SNIa, BAO, $H(z)$ for
galaxies and AP test).

In the model under consideration the energy
density of dark matter is a growing function with the cosmological
time on the cost of dark energy sector. In the basic formulas on
$H^2(z)$ some additional terms appear related with the diffusion
process itself. These contributions can be interpreted as the
running Lambda term ($\bar\Omega_{\gamma,0}\ne 0$) and a
correction to the standard scaling law $\propto a^{-3}$ for dark
matter. At the present epoch the value of the density parameter
related with the dark matter correction is about $1\%$ of total
energy budget.

In the first model it is assumed dark matter in the form of dust.
The estimated values of the model parameters are comparable with
the parameters for the $\Lambda$CDM model and the value of reduced
chi-square of this model is 0.187767. We also studied the second
model with diffusion in a late time approximation: $ma \to \infty$. 
The value of density parameter of $\Omega_{\gamma,0}$
related with diffusion is equal 0.0106. In this case the value of
reduced chi-square is 0.188201. For comparison, the value of
reduced chi-square of the $\Lambda$CDM model is 0.187483.

The value of $\Delta\text{BIC}=\text{BIC}_i-\text{BIC}_{\Lambda\text{CDM}}$ for the first model is 6.421 and for the second model is equal 6.690. While the evidence is strong in favor of the $\Lambda$CDM model in comparison to our model, our model cannot be rejected based on our statistical analysis.

\acknowledgments{The work was supported by the grant NCN DEC-2013/09/B/ST2/03455. The authors thank the anonymous referee for valuable remarks and comments.}

\providecommand{\href}[2]{#2}\begingroup\raggedright\endgroup
\end{document}